# Inclusive AR/VR: Accessibility Barriers for Immersive Technologies

Chris Creed[1], Maadh Al-Kalbani[1], Arthur Theil[1], Sayan Sarcar[1] and Ian Williams[1]

[1] DMT Lab, Birmingham City University

**Abstract.** Augmented and virtual reality (AR/VR) hold significant potential to transform how we communicate, collaborate, and interact with others. However, there has been a lack of work to date investigating accessibility barriers in relation to immersive technologies for people with disabilities. To address current gaps in knowledge, we led two multidisciplinary Sandpits with key stakeholders (including academic researchers, AR/VR industry specialists, people with lived experience of disability, assistive technologists, and representatives from national charities and special needs colleges) to collaboratively explore and identify existing challenges with AR and VR experiences. We present key themes that emerged from Sandpit activities and map out the interaction barriers identified across a spectrum of impairments (including physical, cognitive, visual, and auditory disabilities). We conclude with recommendations for future work addressing the challenges highlighted to support the development of more inclusive AR and VR experiences.

## 1  Introduction

Immersive experiences such as augmented and virtual reality (AR/VR) hold significant potential to address the digital divide for people with disabilities [1, 2] through presenting new opportunities to support social interaction in virtual environments. With the ever-expanding interface that immersive environments have on everyday lives, and the multi-user, socially collaborative and pervasive concepts proposed within the Metaverse, immersive technology will increasingly become part of how we all work, play, communicate, interact and collaborate [3].

Over recent years research has explored immersive technology and virtual environments for users with a range of impairments with the aim of illustrating the potential of this emerging technology [4]. This has led to a range of work across core domains such as assistive AR/VR hardware and software systems. Notable work addressing novel hardware infrastructures includes assistive systems to support users with motor impairments [5], non-invasive cognitive control devices for gaming [6], and supporting user independence and control in smart home environments [7]. Software-based AR/VR solutions have focused on a range of application areas (e.g. assistive visual systems [8, 9] and supportive audio applications [10]) with an emphasis on presenting accessible user experiences. A body of research has also aimed to illustrate and define the psychological and physiological barriers users with impairments face when using AR/VR systems. Specific attention has focused around evaluating the challenges and differences in quantifiable psychological and physiological measures, namely in user sense of presence in the immersive environment [11, 12, 13], in responses to negative experiences (e.g. cybersickness [14] and latency [15]), in adaptations to user motion (e.g. in locomotion [16]) and gait variability [15, 12, 13]. These studies are highlighting significant challenges in current and emergent technology design which is clearly illustrating the implicit lack of design considerations for users with a spectrum of impairments.

Moreover, whilst studies are demonstrating the future potential for AR/VR systems, limited work has sought to address the fundamental accessibility barriers associated with readily available and emerging consumer level AR/VR hardware and software platforms. The development of the Metaverse will promote substantial future uptake of AR/VR technology across the wider population and unless addressed the accessibility barriers inherent in this technology will present significant issues for people with disabilities. This will further exacerbate the digital divide, leading to exclusion from new social environments and a restriction in working opportunities and access to future systems. Recent work has therefore proposed a call to action [14] to address the lack of activity in this area. In particular, the call highlights the need for designers, researchers, developers, and users of AR/VR technology to place an emphasis on disability within all facets of their work, including using inclusive imagery within marketing of devices, in widening diversity and representation in participant recruitment, in developing stronger relationships with affinity groups, and in reporting different perspectives and accurate diversity data within studies [14]. An essential element in supporting this call is a clearer understanding of existing barriers of the technology for people with disabilities to help inform where future research needs to focus in the short-medium term to help accelerate impact in this area.

To address the limited work in this area, we led two full day "Sandpits" to explore the challenges and unique issues associated with AR/VR experiences. These events were multidisciplinary in nature and comprised of key expert stakeholders including people with lived experiences of disability (across the spectrum of physical, visual, auditory, and cognitive impairments), national charity representatives, assistive technologists, AR and VR industry specialists, special needs schools and colleges, and academic researchers specialising in Human-Computer Interaction (HCI), AR/VR, and accessibility. A thorough analysis was conducted across the two Sandpits where 49 current barriers were identified and grouped into four core themes: Software Usability, Hardware Usability, Ethics, and Collaboration and Interaction. This work therefore presents a significant contribution through mapping out of the key existing challenges currently experienced by people with physical, visual, auditory, and cognitive impairments when using AR/VR technologies. Furthermore, our work provides a platform for academic, industry, and other invested organisations to start addressing the core challenges identified to ensure people with disabilities are not excluded from immersive experiences that can support future opportunities for inclusive communication and social interaction.

## 2 Related Work

With the recent advances in AR/VR there is now increased activity in research to determine the future roadmap for this technology in shaping the way we all socialise, interface, collaborate, work and play [17]. Complementary to this, the current availability of commercial AR/VR devices such as Microsoft HoloLens 2, Meta Quest 2, HTC Vive, Magic Leap, and future devices in the pipeline from technology giants such as Apple, Meta, and Microsoft, mean that AR/VR technology is becoming increasingly more accessible to a wider population.

Over many years research has presented and evaluated the future potential benefits of AR/VR technology. This has focused on novel applications and processes where immersive technology can enrich our everyday lives [18, 19]. However, while work has addressed the potential of this emergent technology, less work has focused on the potential of AR and VR to enrich the lives of people with disabilities. Where research exists, there is a primary focus on using commercially available devices to present findings which develop and evaluate assistive software and assistive hardware solutions. While this presents a corpus of knowledge showcasing the potential positive value in AR/VR technology, less work has focused on the fundamental accessibility barriers inherent in commercial AR/VR devices, specifically when considering users with a spectrum of impairments, namely physical, cognitive (neurodiverse), hearing, and visual impairments.

This section presents the current state of art in AR/VR research covering this spectrum of impairments. We present an overview of the key developments to date and highlight the range of activity in AR/VR software and hardware solutions. We additionally illustrate the lack of work which has focused on the core accessibility barriers of commercial AR/VR devices, which if addressed, will support AR/VR in delivering on the promise to create an inclusive and accessible immersive future for all users.

*Assistive AR/VR Developments*
A range of research studies have illustrated the potential benefits AR/VR software and novel hardware systems can have for disabled users. For users with physical impairments, research has explored applications supporting rehabilitation [20, 21, 22, 23], physiotherapy [24], biomechanical movement [25] and for stabilising physical involuntary motion (i.e. hand tremors) [26]. Additionally, immersive technology research has explored interactive gaming systems and interfaces for assisting wheelchair users [27, 28, 29]. For users with visual impairments, research has illustrated how immersive technologies can be used as visual aids to support environmental awareness and promote sensory substitution [45]. Furthermore, work has focused on novel interaction methods for expanding users' spatial awareness [46, 47, 48] and for developing novel user interaction techniques which combine object localisation and spatial audio [49, 50] and echolocation [51]. Haptic interactions has also been used with immersive technology for users with visual impairments in novel interfaces to support sensory substitution [52, 53, 54]. Moreover, literature has explored the use of feed-forward techniques for immersive technology to support users with visual impairments in virtual interactions [55, 47, 56].

For users with neurodiversity and those living with cognitive impairments, research has focused on the potential of AR/VR for delivering innovative solutions that offer the ability to manipulate and specifically target cognitive, sensory, interpersonal, and motor processes that contribute to atypical developmental trajectories [30]. This potential for immersive experiences as supportive tools has been largely explored in literature across a spectrum of neurodiverse conditions, notably Dyslexia [31, 32], Dysgraphia (i.e. difficulties in writing) [33, 34], Dyscalculia (difficulties in performing arithmetical calculations) [35]. Additionally, AR and VR interventions have been developed for users living with Mild Cognitive Impairments (MCI), Dementia and age-related impairments [36, 37]. For intellectual and developmental disabilities, research has focused on developing assistive tools for users living with Autism Spectrum Disorder (ASD) [38, 39], Attention-Deficit Hyperactivity Disorder (ADHD) [40, 41], Obsessive Compulsive Disorder (OCD) [42, 43] and Dyspraxia (i.e. difficulties in performing coordinated movements) [44].

Research has also focused on exploring new accessible solutions for people who are Deaf and Hard of Hearing (DHH) [57] with a key focus around facilitating communication and social interaction. In particular, work has explored AR/VR systems for providing visual support to enhance communication, notably for representing conversations as speech bubbles [58] and using technologies such as Automatic Speech Recognition (ASR) to support visual augmentation in social conversations [59]. ASR has also been applied within AR/VR digital human representation and for collaboration with avatar representations [60], as well as for creating interactive narrative and educational textbooks [61, 62]. Furthermore, work has explored to what extent AR software solutions can support wider social environment representation and social communication support, namely in improving vocal pronunciation and language learning [63], or for improving parental and child communications [64]. Finally, AR/VR assistive software research has reported on novel AR/VR tools for visualising sign language communication [65] and for supporting enriched environment object localisation [66].

While clearly demonstrating how AR/VR technology can offer valuable multi-faceted benefits for users with impairments, we still lack a clear and thorough understanding of the underlying barriers that can be experienced when first accessing or using immersive technologies.

*AR/VR Accessibility Barriers*
Initial research has started to explore the barriers and challenges experienced when using immersive technologies. This has largely focused on challenges faced by users with physical impairments [67, 68, 69, 70] including social support and communication barriers [71, 72, 73]. Whilst this initial work has been crucial in identifying initial challenges, much of the emphasis has been on challenges for wheelchair users and does not yet cover the full spectrum of physical impairment. Complementary to this work, research has started to explore the barriers for neurodiverse users. In particular, work has focused on the challenges faced by users with ASD in terms of sensory inputs [74] and barriers relating to the lack of text customisation, workplace distractions, social interaction challenges, nausea and cybersickness, confusion, motion sickness, eye strain and anxiety [75, 76]. Research has also highlighted barriers such as incompatibility of HMDs with physical accessibility aids (e.g. glasses), discomfort of VR headsets, agitation, unintentional damage of VR headsets in aged care settings, and misinterpretation of reality (e.g. users believing VR is real) [76].

Whilst there has been some initial work exploring the barriers for people with physical impairments or those who are neurodiverse, there has been much less research in terms of barriers for users with visual and auditory impairments. For instance, research has highlighted that current virtual reality systems do not currently support users who are blind or experience low vision [56, 77] and emphasised the importance of adaptability within interactive experiences [78], although there currently remain significant gaps in our understanding. Further work is now essential across all forms of impairment to supplement the initial work completed to date and to more thoroughly scope out the core accessibility barriers that need to be addressed to support the development of inclusive AR and VR experiences for all.

## 3    Methodology

To address the limited work to date around understanding the current accessibility barriers and challenges associated with AR and VR experiences, we led two multidisciplinary Sandpits including academic researchers (specialising in areas such as HCI, immersive technologies, and accessibility), AR/VR industry specialists (e.g. GlaxoSmitheKline, UltraLeap, etc.) people with lived experience of disability (across a range of impairments), and representatives from national charities (e.g. RNIB, Leonard Cheshire, Anne Sullivan Foundation, Everyone Can, Royal Association for Deaf People), special needs schools and colleges (e.g. Bridge College, Treloar School and College), and assistive technologists. Participatory and user centric design methods allow a deeper understanding of user needs, strengths and experiences [82]. Additionally, user centric methods also mediate constant feedback and engagement with relevant stakeholder across all stages of development, to ensure development of an end-product that is usable and sensitive of user needs and characteristics [83]. This approach was taken as we felt it was crucial to get a wide range of different perspectives to deeply understand the spectrum of accessibility challenges associated with AR and VR technologies.

Both Sandpits were full day events held remotely on Microsoft Teams where participants were asked to discuss a range of topics related to existing barriers in using AR and VR (Institutional Review Board approval was obtained for the project). The first Sandpit was held on 11th November 2021 and involved 22 participants (including 7 academics from the organising research team) who had some previous awareness and experience with immersive technologies. Participants were contacted via email prior to the day to provide them with an information sheet highlighting key activities and responsibilities during the Sandpit. Participants were also requested to provide their consent prior to attending on the day and to request any access requirements. The day began with an introduction from the research team highlighting the motivation behind the project and the planned activities for the day. In particular, it was highlighted that the emphasis of day was on discussing AR and VR in context on head-mounted displays (HMDs), as opposed to mobile applications of the technology. This decision was taken as accessibility challenges associated with mobile devices are relatively well-understood, although less work has focused on the barriers related to wearable immersive experiences.

Following the introduction, participants were divided into three separate groups comprised of seven participants (including two facilitators from the research team) to explore and discuss existing accessibility barriers associated with immersive technologies. In selecting participants for different groups, we aimed to achieve a balance across academic researchers, people with disabilities, charity representatives, and assistive technologies to ensure a range of perspectives could be presented. In the morning activities, participants within each group were initially asked to share any barriers or limitations with AR and VR from their own experiences (primarily as an "ice-breaker" task). This was followed by an open group discussion exploring the barriers and challenges associated with immersive technologies. In the afternoon, the discussion focused on requirements for developing more inclusive AR/VR experiences (linked to the barriers identified in the morning activities). All groups reconvened at the end of the day with lead facilitators summarising key points and findings from the discussions. Participants were paid £250 for their time and contribution.

This discussion provided a wide range of insights (detailed in Section 4) and helped to inform the design of the second Sandpit. In particular, to supplement the key themes that emerged, a decision was taken to structure the next event around different forms of impairment to help facilitate a deeper discussion around specific barriers and challenges. The second Sandpit therefore followed the same underlying procedure, although participants were instead divided into four separate groups (comprised of 7-9 participants) focused on physical, cognitive, visual, and auditory impairments. The event was held on 21st January 2022 and involved 33 participants (including 9 research team members) with 14 attendees self-disclosing some form of disability (across the spectrum of physical, cognitive, visual, and auditory impairments). To support accessibility on the day, two BSL interpreters also attended and could be "pinned" to the display by individual participants - guidance was also provided around how to enable auto-captioning features within Teams for those who required this feature.

In terms of attendees, we were keen to encourage participation from people with different lived experiences of disability who have varying degrees of experience with AR and VR (including those who have not been able to use these technologies yet, but have a desire to do so). To capture the diversity of experience, participants (excluding the research team) were asked to specify their level of experience with immersive technologies at the start of the morning activities (from a scale of 1-5 – where 1 relates to "No Experience" and 5 "Very Experienced"). Scores covered the full-scale range with 9 participants providing a score of three or above, 7 participants choosing a rating of two, and the reminder highlighting that they have little or no significant experience.

The morning session then focused on initially providing a definition of immersive technologies to ensure all participants had a shared understanding (given that some participants had less experience in using the technology), followed by a group discussion around exploring barriers around inclusive AR/VR experiences. The afternoon session focused on the Metaverse and the challenges associated with inclusive communication and social interaction within collaborative and shared virtual environments. Group discussions across both Sandpits were recorded for later analysis with participants' consent. Participants were also again paid £250 for their time and contribution.

## 4    Results

A thematic analysis was conducted across all Sandpit video recordings where key points were coded to identify initial themes and barriers - these themes were then iteratively refined and shaped through exploring the relationship between different barriers. The key barriers and themes captured are detailed below and are structured in relation to different forms of impairment (i.e. physical, cognitive, visual, and auditory) and the four high-level categories derived through the thematic analysis – Software Usability, Hardware Usability, Ethics, and Collaboration and Interaction.

**Physical Impairments**

| Software Usability |
|---|
| **Involuntary movements:** challenges around environment navigation using AR/VR headsets and controllers for users with involuntary limb or eye movements |
| **Fatigue (physical, mental, temporal):** concerns around fatigue associated with the use immersive experiences and its impact on existing physical conditions |
| **Real world physical awareness and proprioception:** concerns around the risks associated with losing track of physicality, balance, and perception of limbs in fully immersive environments |
| **Lack of personalization and dynamic mapping of user reality:** current AR/VR systems do not consider unique user characteristics and offer no personalization for users that may have specific software and hardware needs |
| **Hardware Usability** |
| **Limited physical movement:** challenges around wearing AR/VR devices securely and accurately by users with limited physical movements |
| **Facilitation of physical use\interface use:** challenges associated with current lack of AR/VR support on wearing devices, navigating environments, menus, and buttons during (and prior to) AR/VR use |
| **Lack of compatibility and integration with existing mobility aids:** concerns around suppression of communication due to lack of access to physical communication and assistive aids while using AR/VR |
| **Physical Device Form Factor, Design, Ergonomics:** concerns around usability, weight and comfort of AR/VR HMDs and controllers |
| **Ethics** |
| **Psychological, mental, and emotional impact:** lack of clarity around the potential psychological, emotional, and mental impact of AR/VR on users living with physical impairments |
| **Unethical design, unconsidered and unbounded use:** concerns around inherited problems from social media sites such as discrimination, cyberbullying and excluding users in collaborative virtual spaces such as the Metaverse |
| **Choice and physical representation:** concerns around user representation using avatars and potential lack of measures for sharing identities in immersive environments |
| **Collaboration and Interaction** |
| **Hand control/manual/bimanual and limb interactions:** interaction challenges using AR/VR headsets and controllers that require user can move limbs with great dexterity and speed |

*Table 1: Key barriers to using AR/VR technologies for users living with physical impairments across four themes (software usability, hardware usability, ethics, and collaboration/interaction)*

*Software Usability*

For users with lived experiences of physical impairments, several software usability barriers were identified (Table 1). For instance, involuntary body movements present a challenge to use AR/VR input devices as methods of interaction and collaboration in AR/VR experiences. Participants indicated that this barrier is also applicable for users with involuntary eye movements and is particularly problematic in AR/VR systems that depend on constant targeted eye movements for navigation, interaction, and collaboration. Physical, mental, and temporal fatigue during prolonged use also presents a key barrier to using AR/VR. This extends beyond the well-known potential cybersickness side-effects associated with AR/VR input devices as it can cause aching for users with movement disorders, and potentially worsen certain symptoms or conditions for users living with physical impairments. Users indicated worsening of symptoms was also applicable to certain mental conditions such as psychosis that can be negatively impacted by VR/AR exposure.

    Lack of real-world awareness (i.e. balance) is another barrier highlighted by participants that leads to losing track of physicality, physical aids, and balance in immersive environments primarily due to full immersion of VR systems. The impact of complete disengagement from the physical world can be particularly detrimental for users with physical impairments as it elevates physical and mental injury risks, and can lead to users disengaging with immersive experiences altogether. Closely related to the lack of real-world awareness barrier is proprioception (i.e. body awareness) that presents a significant challenge for users who are not able to localise different body parts (e.g. hands or legs) when fully immersed in VR environments. Additionally, the lack of

customisation and dynamic mapping of user reality was also highlighted as key barrier to using AR/VR.

*Hardware Usability*

Discussions highlighted how users living with limited physical movements can find it challenging to wear AR/VR headsets securely and accurately thus presenting a key entry level barrier to using AR/VR technologies. Likewise using controllers with the required levels of motion trajectories and dexterity becomes a challenge and often leads users to stop using immersive experiences. Lack of support during AR/VR use was also highlighted as a barrier. In particular, mediating external assistance during AR/VR interactions on how to wear devices, navigate environments, interact with menus and buttons is important for users living with physical impairments who may not be AR/VR literate or simply need physical and emotional support during use. Users also indicated that they would ideally prefer facilitators to see what the user is experiencing in the immersive environment to provide the best support during use of AR/VR.

Lack of compatibility and integration with existing physical mobility aids is another key barrier that leads users to feeling less confident when using AR/VR systems. For example, users on crutches or wheelchairs find movements in AR/VR challenging and restrictive due to multi-tasking and the speed of interaction required. Participants also highlighted that it is currently challenging to use AR/VR devices for users that wear accessibility aids (e.g. glasses, canes, hearing aids). The lack of compatibility with physical aids in current AR/VR systems was highlighted as being particularly frustrating by participants, and usually leads users to stop engaging with the technology. Participants indicated that current AR/VR systems offer no personalization for users that may have specific hardware (and software) needs and highlight the current lack of bespoke and tailored AR/VR systems for users with physical impairments. Additionally, participants also highlight discomfort of current AR/VR headsets as a key barrier to sustained engagement in immersive environments.

*Ethics*

Several ethical concerns and barriers were highlighted during discussions – for example, participants highlighted the lack of clarity around the potential psychological, emotional, and mental impact of AR/VR on users living with physical impairments as a barrier and stressed the need for more research in this area. Participants also emphasised that using AR/VR tools to alter realities can potentially be damaging to users as the concept of reality can be very different for users living with disabilities than non-disabled users. Additionally, participants highlighted that manipulation of realities using AR/VR and how users experience them can be a disorientating experience that makes it emotionally challenging for some users living with disabilities to leave immersive environments.

Unethical design, unconsidered, and unbounded use were also highlighted as key barriers to using AR/VR - participants were generally concerned regarding the replication of inherited problems from social media sites such as discrimination, cyberbullying and the unmonitored ability to exclude users in AR/VR and highlighted this as an entry barrier to consider using the technologies. Like inherited problems from social media platforms, the unknown impact of algorithms and targeted adverts in collaborative and social AR/VR environments also presents a key barrier that can potentially become magnified in the developing concept of the Metaverse. Participants also raised the question on what measures are currently in place to prepare users to step into VR/AR environments and how to interact within them in a safe way, and pointed out that it would be desirable and logical to train users on AR/VR worlds just like they are trained to deal with real world scenarios, especially if collaborative immersive environments (such as the Metaverse) are to become platforms for future work, collaboration and social interaction.

Additionally, appearance choice and physical representation concerns were raised by participants as potential barriers to using AR/VR. Users indicated that providing more choice (or complete freedom of choice) of avatars is important for people who struggle with body image and how other users perceive it, as well as underlining the need for avatar choice to be treated with care if sharing identities is required in collaborative AR/VR environments such as the Metaverse.

*Collaboration and Interaction*

With regards to collaboration and interaction in immersive environments, participants have highlighted difficulties in head, manual, bimanual and limb interactions as a key barrier to using AR/VR. Participants indicated that current devices or interactions assume the user can move limbs and other body parts (e.g. head, eyes) with great dexterity which is not the case for many users living with physical impairments. Likewise, participants emphasised that current AR/VR controllers do not accommodate for different needs of users living with physical impairments as they are usually designed in a one-size-fits-all manner that lacks flexibility and customisation. These barriers present significant challenges for users living with physical impairments to make the most out of virtual collaborative spaces, and significantly hinder user experience in immersive environments.

**Visual Impairments**

| Software Usability |
|---|
| **Lack of binaural audio:** barriers associated with current lack of integrated audio descriptions for environment description, navigation, and interaction |
| **Voice activation and accessible menus:** challenges associated with inaccessible menus in AR/VR systems |
| **Lack of built-in accessibility features:** concerns around lack of built-in accessibility features for users living with visual impairments |
| **Sensory and/or information overload:** usability challenges associated with sensory, information and cognitive overload in immersive environments |
| **Lack of customisation:** challenges linked to inability to adjust different features of immersive environments based on the disability and reality of users |
| **Hardware Usability** |
| **Lack of standardisation of the headset system:** concerns around lack of standardisation, guidelines, and protocols for developing consumer products for users living with visual impairments |
| **Haptics:** usability challenges associated with haptics (e.g., cost, setting up time and comfort) |
| **Ethics** |
| **Representation of visually impaired users:** concerns around the lack of clarity on how visually impaired people will be represented in shared virtual spaces such as the Metaverse |
| **Darker side of humanity:** concerns around inherit societal issues experienced in current social and collaborative platforms (e.g.,cyberbullying, harassment, exclusion of disabled users) |
| **Constant adding of accessibility features as add-ons:** concerns around extra costs associated with accessibility add-ons for AR/VR systems |
| **Collaboration and Interaction** |
| **Environment and user awareness:** challenges associated with lack of environment and user awareness in collaborative immersive environments |
| **Entry point barriers (prior to collaboration):** entry barriers that occur prior to collaboration (e.g. users living with visual impairments not being able to find the button that switches AR/VR headsets on) |

*Table 2: Key barriers to using AR/VR technologies for users living with visual impairments across four themes (software usability, hardware usability, ethics, and collaboration/interaction)*

*Software Usability*
The lack of binaural audio in current AR/VR systems was highlighted as a key barrier that presents challenges in navigation and interaction in immersive environments (Table 2). Participants also emphasised that immersive environments currently lack audio description and stressed the need for more research in three key areas – namely, audio being integrated as part of the system, audio description of the environment and audio for navigating the environment. Voice activation and accessible menu design was another key barrier highlighted by participants – in particular, attendees felt that voice activation mechanisms in current AR/VR systems are not sufficiently advanced and do not provide ways of interacting with the system (e.g. using phrases such as "link, bring up the menu, zoom in, zoom out"). This presents a significant entry level barrier for users living with visual impairments as the point of entry to the system is inaccessible in current applications. Participants argued that while research around simpler and more usable interaction techniques in AR/VR environments is appreciated, the definition of simplicity for users living with visual impairments is different and, in many cases, revolves around being able to find menus in immersive environments to begin with before interaction takes place.

Participants also highlighted the lack of built-in accessibility features as a key barrier to using AR/VR technologies and urged leading industry manufacturers of state-of-the-art headsets to develop built-in accessibility features to improve accessibility in AR/VR environments in the long run. Even though users acknowledged that there are out of the box solutions that can be used as add-ons, opting for these solutions is not desirable by users living with visual impairments as they would ideally prefer these separate features and tools to be fully integrated in AR/VR hardware and software at no extra cost. Sensory and/or information overload during use of AR/VR also presents a barrier to using these technologies for users with visual impairments. Brightness in particular was highlighted by users to be problematic during use, especially if use is prolonged during collaborative scenarios in immersive environments. Furthermore, cognitive overload was highlighted as a risk, especially when users are required to interpret non-verbal communications as well as interactions and environment navigation in AR/VR.

Additionally, lack of customisation in current AR/VR systems and the inability to adjust different features of immersive environments based on the disability and reality of the user (e.g. adjust sensory information, accessibility features) was highlighted as posing a key barrier for users with visual impairments.

*Hardware Usability*
In terms of hardware usability, participants called for standardisation of the headset system for users with visual impairments and argued that there are currently not enough guidelines and protocols in place for developing inclusive consumer products. Participants indicated that users living with visual impairments would ideally prefer dedicated AR/VR headsets, with disabled people at the centre of development and design, where users can change and adapt settings/preferences and can plug into other systems and experiences. Recent progress in the use of haptics to improve accessibility in AR/VR environments is appreciated by users with visual impairments, however participants indicated that haptic interaction remains unstandardised and presents its own set of unique barriers such as cost, setting up time and user comfort.

*Ethics*
Users voiced their concern around the lack of clarity on how visually impaired people will be represented in collaborative immersive environments such as the Metaverse, and how they can use it to interact with others. Participants stressed the importance of representation of visual disabilities to mitigate this barrier and avoid conflict or misunderstanding of behaviours within collaborative AR/VR environments or the developing Metaverse. The "darker side of humanity" as described by participants refers to the inherit societal issues experienced in current social and collaborative platforms (e.g. cyberbullying, harassment, exclusion of disabled users or "misfits") and presents another key ethical barrier to using AR/VR for user living with visual impairments.

The constant trend of adding of accessibility features as add-ons also remains a significant barrier to using AR/VR for users living with visual impairments. Participants indicated that people with disabilities are not currently entitled to get accessibility features at the same time as non-disabled users or consumers, and they attribute this trend to three core reasons: high costs, lack of inclusion of users with impairments in the designing and development stages of AR/VR hardware/software, and lack of knowledge around the creators of AR/VR technologies with many apparent missing voices from disabled communities.

*Collaboration and Interaction*

In terms of collaboration and interaction in immersive environments, participants highlighted lack of awareness to surroundings and users in immersive environments as a key barrier in collaborative settings such as the Metaverse. Additionally, participants highlighted the significance of entry barriers that occur prior to collaboration (e.g. users with visual impairments not being able to find the button that switches AR/VR headsets on), with one participant stating, "if you cannot see the environment, then how can you start using it and collaborate within it?".

**Neurodiversity / Cognitive Impairments**

| Software Usability |
|---|
| **Detachment from the real environment:** concerns around the implications of complete immersion on physical and mental wellbeing |
| **Cybersickness and usage aftereffects:** challenges around cybersickness and usage aftereffects of AR/VR technologies (e.g. motion and simulator sickness, disorientation, lagged feeling of immersion after exposure) |
| **Sudden unexpected changes in immersive experiences:** uncertainty and stress associated with sudden unknown changes in immersive environments (e.g. changes in brightness, avatars, movements) |
| **Sensory and information overload:** usability challenges associated with sensory, information and cognitive overload in immersive environments that are not customisable |
| **Hardware Usability** |
| **Discomfort of HMDs:** challenges around discomfort of AR/VR devices in terms of weight, tightness, pain with prolonged use and incompatibility with physical assistive aids |
| **Physical or mental injuries or stress:** concerns around physical or mental injury risks when fully immersed in AR/VR environments |
| **Lack of support and training:** difficulties around setting up AR/VR devices where assistance is usually needed to setup, wear and adjust these devices in a way that is comfortable for neurodiverse users |
| **Ethics** |
| **Darkside of humanity:** concerns around inherited societal problems in current virtual collaborative and social interaction environments (e.g. abuse, cyberbullying and exclusion of users) |
| **Peer pressure and addiction:** concerns around replication of negative traits of current social interaction platforms (i.e. per pressure and excessive use) in future collaborative environments such as the Metaverse |
| **Protection to vulnerable users:** barriers associated with current lack on measures that will be taken to protect and support vulnerable users in shared virtual spaces like the Metaverse |
| **Impact of hyper-realism:** concerns around the impact of hyper-realism on neurodiverse users |
| **Physical isolation and inability to separate reality and virtual reality:** challenges around physical isolation and inability to separate reality from virtuality for neurodiverse users that have different standards of reality |
| **Collaboration and Interaction** |
| **Input and hand-eye coordination difficulties:** challenges around input and hand-eye coordination in interactive and collaborative scenarios where multitasking processing of more than one stimulus is required |

*Table 3: Key barriers to using AR/VR technologies for neurodiverse users across four themes (software usability, hardware usability, ethics, and collaboration/interaction)*

*Software Usability*

For users living with neurodiverse and cognitive impairments, participants identified several software usability barriers (Table 3). Complete detachment from the real environment and real-world physicality due to full immersion in immersive environments was highlighted as a key barrier that causes a great deal of stress to neurodiverse users. Cybersickness and usage aftereffects of AR/VR technologies were also highlighted as key barriers. Participants shared that they usually experienced motion and simulator sickness when using AR/VR systems, in addition to some usage aftereffects such as disorientation, feeling immersed in the system after exposure and difficulties in adjusting to the real environment after AR/VR exposure. Users also underlined sudden unexpected changes in immersive experiences as a key barrier (e.g. changes in brightness, movements, avatars) that causes stress and uncertainty about the generated AR/VR environment during use and leads users to stop using the technology overall. Furthermore, sensory and information overload was highlighted as an important barrier to using AR/VR technologies for neurodiverse users. Coupled with the lack of customisation available in current AR/VR systems (e.g. adjust sensory information, brightness etc.), the implications of this barrier on AR/VR adoption by neurodiverse users were highlighted as being particularly significant.

*Hardware Usability*

Current AR/VR devices present several barriers for neurodiverse users. Discomfort of AR/VR devices in terms of weight, tightness, pain with prolonged use and incompatibility with physical assistive aids (e.g. glasses) was a key barrier highlighted by participants. Participants further indicated that in-person verbal instructions are usually needed to setup AR/VR hardware, especially when used in disabled communities and homecare settings where AR/VR literacy is low. Physical or mental injury stress is another barrier faced by neurodiverse users that often worry about physical injury or simply falling when fully immersed in AR/VR environments. Participants also indicated that a carer or family member is usually needed to be present in a supportive role during AR/VR exposure to mitigate the stress faced by neurodiverse users and intervene when needed if the experience becomes risky to the user.

Additionally, participants highlighted the lack of support and training as a key barrier to using AR/VR. Attendees emphasised that they face difficulties in setting up AR/VR devices and assistance is usually needed to setup, wear and adjust these devices in a way that is comfortable for neurodiverse users. Participants further added that support is also needed in learning about AR/VR software and development with one participant stating that AR/VR technologies are "not easy to learn", thus even if an impairment community, care home or a special needs college get access to the latest AR/VR devices and hardware, there still remains the problem of the digital divide in terms of AR/VR literacy between able bodied and impaired communities that limits the potential positive impact AR/VR can have for neurodiverse users.

*Ethics*

Similar to the point raised by participants in relation to visual impairments, neurodiverse users shared a common concern around the "darker side of humanity" as an entry barrier to consider using immersive technologies. Abuse, cyberbullying and exclusion of users in virtual collaborative and social interaction environments were some of the concerns shared by participants. Neurodiverse users also highlighted peer pressure and addiction (i.e. excessive use) that could potentially be inherited from current social interaction and collaboration platforms as barriers to using environments such as the Metaverse. Lack of protection to vulnerable users was also highlighted as a barrier for neurodiverse users with participants sharing common concerns around the lack on measures around protecting and supporting vulnerable users in shared virtual spaces like the Metaverse.

The impact of hyperrealism on the physical and mental wellbeing of neurodiverse users, a concept that is promised by shared virtual spaces and the Metaverse, was another key barrier highlighted. Participants stressed the need for more research to fully understand the impact of hyperrealism and argued that research and development efforts that attempt to improve accessibility in AR/VR by increasing realism are not necessarily effective or user centric, with one participant stating, "more realism does not equal more accessibility".

Additionally, physical isolation and inability to separate reality from virtuality was another key barrier highlighted by participants that is closely related to hyperrealism and long exposure to AR/VR. Participants shared that neurodiverse users may have different standards and definitions of reality (e.g. some users may perceive virtual reality to be a better environment than their real world). It was felt that the risk implications of this particular barrier on neurodiverse users could be significant in hyper realistic AR/VR environments.

*Collaboration and Interaction*

In terms of collaboration and interactions, participants highlighted input and hand-eye coordination difficulties as a key barrier for neurodiverse users. Participants indicated that multitasking processing of more than one stimulus in collaborative settings is overwhelming for neurodiverse users. This barrier can potentially be magnified if neurodiverse users are expected to interact with more than one person and potentially more than one avatar in shared collaborative settings in AR/VR. Participants also linked this barrier to sudden changes in immersive environments and stressed the need to consider managing this carefully to ensure neurodiverse users can safely collaborate and interact with other users in immersive environments.

**Auditory Impairments**

| |
|---|
| **Software Usability** |
| **Friction initial access:** entry level barriers that occur prior to interaction (e.g. AR/VR literacy, technology acceptance) |
| **Lack of clarity in sounds and instructions in audio format:** concerns around poor sound quality and lack of visual prompts for users with hearing impairments |
| **Lack of standardisation in text presentation:** challenges associated with text presentation in AR/VR environments that does not fully describe and reflect immersive environments in terms of context and emotions |
| **Difficulty pinpointing locations and environment navigation:** concerns around head motion tracking navigation in immersive environments where sound sources are unknown or unclear |
| **Hardware Usability** |
| **Space for hearing aid:** concerns around limited physical space for hearing aids in current bulky AR/VR devices |
| **Compatibility and integration with existing technologies and assistive devices:** challenges associated with incompatibility of current AR/VR devices with different coping methods used by people with hearing impairments (e.g. lip reading, assistive devices, audio transcription or sign language) |
| **Haptics:** concerns around long setup times, high costs, and the impact of haptics for prolonged times on users with hearing impairments |
| **Lack of customization:** concerns around current lack of customisable AR/VR devices that are compatible with existing assistive technologies and tools used by users with auditory impairments |
| **Ethics** |
| **Deaf communities can be closed:** barriers around the closed nature of Deaf communities and the potential implications on AR/VR adoption and usage |
| **Collaboration and Interaction** |
| **Inability to use sign languages:** challenges faced by users that are unable to use sign language (due sign language not being the first language or due to low camera and visual quality) |
| **Lack of synchronization in conversations:** challenges around lack of synchronisation in conversations in collaborative settings due to slow audio transcription/captioning in current AR/VR systems, using different (and incompatible) assistive methods and not being able to lip read |
| **Poor rendering of avatars does not allow proper lip reading:** challenges in lip reading and sign language interpretation due to poor rendering quality of avatar or/and visual information in immersive environments |

*Table 4: Key barriers to using AR/VR technologies for users living with auditory impairments across four themes (software usability, hardware usability, ethics, and collaboration/interaction)*

*Software Usability*

A range of entry level challenges were highlighted by participants as a barrier faced by users living with hearing impairments including low AR/VR literacy and acceptance of AR/VR technologies by Deaf communities (Table 3). Lack of clarity in sounds and instructions in audio format is also a key barrier for users living with hearing impairments. Participants indicated that current immersive environments provide audio cues for navigation or interaction purposes, however these cues can be missed by users living with hearing impairments and visual prompts is preferable in this case. Additionally, current systems suffer from lack of clarity in sounds, and available solutions for this particular problem are not necessarily effective with one participant stating, "increasing the volume does not solve the clarity of audio problem". This barrier is closely connected to another one identified around difficulties pinpointing locations and navigating immersive environments.

Participants also stated that current head motion tracking navigation can be problematic for users with hearing problems and navigating or interacting with AR/VR environments is challenging if the sound sources are unknown or unclear. Additionally, lack of standardisation in text/subtitle presentation was also highlighted as a key barrier to using AR/VR. Participants acknowledged that current subtitle formats are helpful but highlight that they do not fully describe and reflect immersive environments in terms of context and emotions. This barrier is potentially elevated in immersive environments that lack audio descriptions and descriptions of external sounds.

*Hardware Usability*

Limited space for hearing aids in current bulky AR/VR devices was a key barrier highlighted by participants. Microphones in current AR/VR devices cause feedback when placed above hearing aids, making the experience uncomfortable and leads users to stop using the technology altogether. Lack of compatibility with existing technologies and integration with other assistive devices is another key barrier for this user group. Participants clarified that people living with hearing impairments use different methods to support interactions in different environments (e.g. lip reading, assistive devices, audio transcription or sign language) which may overlap with current accessibility tools. This overlap can also lead to users with hearing impairments being left out of conversations and activities in collaborative settings. Moreover, participants highlighted current lack of integration of tools (including input and output) that are relevant to individuals with hearing problems (e.g. haptics, mouth, and body movements).

Use of haptics in AR/VR systems to support users with hearing impairments was another hardware barrier identified. Participants indicated that haptics are tiresome to use and raised concerns around long setup times, high costs, and the impact of using haptics for prolonged periods on users with hearing impairments. Additionally, participants underlined lack of customisation as a key barrier to using AR/VR technologies and urged headset manufacturers to build devices that are compatible with existing assistive technologies and tools to improve AR/VR accessibility, though participants also acknowledged that developing specific devices for different levels of hearing impairments may not be commercially viable.

*Ethics*

Participants highlighted that Deaf communities are known to be closed with a strong identity and would not seek to fit in or adopt new technologies unless they have been accepted by the community. In particular, when presented with concepts of the Metaverse and shared virtual spaces for collaboration, participants with hearing impairments indicated that they would not "find their tribe" in environments such as the Metaverse if the community has not adopted this form technology first. Participants also highlighted that users with hearing impairments do not have access to AR/VR tools and technologies available to them, and even if they were to be available there remains the problem of AR/VR literacy in Deaf communities.

*Collaboration and Interaction*

Participants highlighted challenges in lip reading and sign language interpretation due to poor rendering quality of avatar or/and visual information in shared immersive environments. This barrier illustrates that even if a user with a hearing impairment is confident and able to use lip reading and sign language, poor quality of relevant visual information can still present a barrier to use AR/VR technologies. Participants further added that critical information can be lost during conversations in collaborative settings if users are required to lip read avatars rather than real people on camera. Participants highlighted "apple's Animoji's" as an example for how current state of the art facial mesh avatars do not yet provide sufficient fidelity in terms of quality to enable accurate lip reading.

## 5   Discussion

This paper presents key barriers faced by users living with a range of impairments (i.e. physical, visual, auditory, and cognitive) when using immersive technologies, utilising a user centric and participatory study design that included participation from users with these impairments and relevant key stakeholders (i.e. charity workers, community representatives, academic and industry experts). Our work presents unique barriers identified by users living with different impairments and relevant stakeholders (see Tables 1-4), and reports on insights and lived experiences shared by users with different impairments around the use of AR/VR technologies. Several themes also align and confirm findings highlighted in related work – for instance, in terms of challenges around setting up immersive systems (including dependency on others), issues around the integration of external assistive tools, potential harm in engaging with experiences that may be inaccessible within real world scenarios, and issues relating to the representation of disability via avatars [67, 68, 69].

Several common themes of barriers and challenges were noted – for instance, entry level barriers that are faced by users prior to starting or experiencing AR/VR experiences were common across different impairments. Challenges such as lack of access to AR/VR technologies and devices in communities representing users living with impairments, low AR/VR literacy and inability to use AR/VR technologies due to current lack of accessibility features present key barriers to AR/VR adoption. Users across groups also highlighted additional entry barriers to using shared virtual spaces and collaborative settings (such as the Metaverse), namely Metaverse literacy, software, hardware, and training requirements that are currently not readily available for users living with impairments. Entry barriers further emphasise the current digital divide for users with impairments [1, 2] which limits access to AR/VR technologies for these user groups.

Users across groups stressed the need of including and embedding disabled users across all stages of AR/VR research and product development (i.e. concept, design, and development). To address this barrier, recent research shows a promising shift towards more inclusive study designs when developing AR/VR for end users living with disabilities [79, 80]. Additionally, several frameworks now provide clear guidelines for designers and developers to establish participatory methods to ensure development of tools that address user needs and capitalise on user strengths [75].

Participants also argued that promising concepts in academic research often stay within universities and do not generally move into production and were critical of academic research studies that use able bodied participants to evaluate AR/VR prototypes and systems that are intended for use by people living with impairments. Insights from participants regarding effectiveness of academic research in addressing AR/VR accessibility needs are also in alignment with findings from a recent review of research methods and practices around the use of emerging technologies to support users living with disabilities [75], which found research studies with users living with disabilities to be mostly exploratory and technologies associated with these studies are usually not tested in long term real-world scenarios. A more recent review on research concerning the impact of AR on adults and children living with ASD also highlight the lack of longitudinal studies as a barrier to fully understand the impact of research studies and associated technological outputs for users living with impairments [81].

Key societal and economical barriers were highlighted that have a significant impact on AR/VR adoption by users with impairments or stakeholders representing and/or caring for them. In particular, participants emphasised the need for financial affordability of AR/VR devices and highlighted the importance of equality of cost where accessibility tools, add-ons and devices would ideally not come at extra costs for users living with impairments as they currently do. Regarding collaboration and shared immersive experiences, our work found that multitasking, or the concept of multitasking in a Metaverse like environment, presents an overwhelming barrier to using AR/VR effectively across impairments. This is caused by the lack of integration of physical accessibility aids used by impaired users within AR/VR devices and systems, that do not presently personalise experiences depending on the accessibility aids used or the physical reality of the user. Additionally, this barrier is also exacerbated through current AR/VR systems typically requiring quick or dexterous limb or controller interactions.

It is essential in terms of future work that the research community urgently starts to address the broad scope of barriers identified to ensure inherent interaction biases around AR and VR can be resolved. This will involve actively exploring and resolving the technical challenges highlighted in terms of hardware and software experiences, as well as considering wider ethical, societal, and economic issues. A collaborative approach also needs to be widely adopted moving forward where people with lived experience of impairments are directly informing and shaping the design of more inclusive immersive platforms at every stage. Wider partnerships with invested stakeholders (i.e. charities, special needs schools/colleges, disability organisations, etc.) can also help to accelerate efforts around developing more accessible AR/VR experiences that are truly effective for people with disabilities.

## 6  Conclusion

Our work contributes a deeper understanding around the range of technical, societal, and economical barriers experienced by people with lived experiences of disability in relation to immersive technologies. The mapping of key challenges to different forms of impairment provides an important platform for the wider research community to start addressing the key accessibility challenges identified. It is crucial moving forward that new approaches and innovative techniques are collaboratively explored and evaluated (i.e. between technical specialists, people with lived experience of impairments, wider stakeholders, etc.) to help facilitate the development of more inclusive AR/VR experiences for all users.

## Acknowledgments

The authors would like to thank Meta Reality Labs for supporting this project through a "Consider Everyone" research grant. We would also like to thank participants for their valuable contributions during the Sandpits.